\newcommand{\diracslash}[1]{#1\llap{/\kern2pt}}

\newcommand{\be}{\begin{equation}}
\newcommand{\ee}{\end{equation}}
\newcommand{\bea}{\begin{eqnarray}}
\newcommand{\eea}{\end{eqnarray}}
\newcommand{\ba}[1]{\begin{array}{#1}}
\newcommand{\ea}{\end{array}}

\newcommand{\bt}{\begin{tabular}}
\newcommand{\et}{\end{tabular}}

\newcommand{\beas}{\begin{eqnarray*}}
\newcommand{\eeas}{\end{eqnarray*}}

\documentclass[preprint,prd,aps,floats,nofootinbib,showpacs,floatfix]{revtex4}
\usepackage{graphicx}
\begin{document}

\title{Charmonium mass in hot asymmetric nuclear matter}
\author{Amruta Mishra}
\email{amruta@physics.iitd.ac.in,mishra@th.physik.uni-frankfurt.de}
\affiliation{Department of Physics, Indian Institute of Technology, Delhi,
Hauz Khas, New Delhi -- 110 016, India}

\author{Arvind Kumar}
\email{iitd.arvind@gmail.com}
\affiliation{Department of Physics, Indian Institute of Technology, Delhi,
Hauz Khas, New Delhi -- 110 016, India}

\begin{abstract}
We calculate the in-medium masses of J/$\psi$ and of the excited states
of charmonium ($\psi$(3686) and $\psi$(3770)) in isospin asymmetric 
nuclear matter at finite temperatures. These mass modifications arise 
due to the interaction of the charmonium states with the gluon condensates 
of QCD, simulated by a scalar dilaton field introduced to incorporate 
the broken scale invariance of QCD within an effective chiral model. 
The change in the mass of J/$\psi$ in the nuclear matter with density 
is seen to be rather small, as has been shown in the literature
by using various approaches, whereas, the masses of the excited states 
of charmonium ($\psi$(3686) and $\psi$(3770)) are seen to have 
considerable drop at high densities. The dependence of the
masses of the charmonium states on the isospin asymmetry has also been 
investigated in the hot nuclear matter and is seen to be appreciable 
at moderate temperatures and high densities. These medium modifications 
of the charmonium states should modify the experimental observables 
arising from the compressed baryonic matter produced in asymmetric 
heavy ion collision experiments in the future facility of FAIR, GSI.

\end{abstract}
\maketitle

\def\bfm#1{\mbox{\boldmath $#1$}}

\section{Introduction}
The study of the in-medium properties of hadrons is an active topic of
research in strong interaction physics, both experimentally and 
theoretically. The topic is of direct relevance in the context of 
heavy ion collision experiments, which probe matter at high 
temperatures and/or densities. The medium modifications of the
hadrons have direct consequences on the experimental observables
from the strongly interacting matter produced in heavy ion collision
experiments. The in-medium properties of the kaons and antikaons
in the nuclear medium are of relevance in neutron star phenomenology
where an attractive interaction of antikaon-nucleon can lead
to antikaon condensation in the interior of the neutron stars.
The medium modifications of kaons and antikaons also show in their 
production and propagation in the heavy ion collision experiments. 
The modifications of the properties of the charm mesons, $D$ and 
$\bar D$ as well as the $J/\psi$ mesons and the excited states of 
charmonium, can have important consequences on the production of open 
charm and the suppression of $J/\psi$ in heavy ion collision experiments.
In high energy heavy ion collision experiments at RHIC as well as LHC,
the suppression of $J/\psi$ can arise from the formation of
the quark-gluon-plasma (QGP) \cite{blaiz, satz}. 

The D (${\rm {\bar D}}$) mesons are made up of one heavy charm quark 
(antiquark) and one light (u or d) antiquark (quark). 
In the QCD sum rule calculations, the mass modifications of 
$D$ ($\bar D$) mesons in the nuclear medium arise due to the 
interactions of light antiquark (quark) present in the $D$($\bar D$) 
mesons with the light quark condensate. There is appreciable change 
in the light quark condensate in the nuclear medium and hence 
$D$ ($\bar D$) meson mass, due to its interaction with 
the light quark condensate, changes appreciably in the hadronic matter.
On the other hand, the charmonium states are made up of a heavy charm quark and 
a charm antiquark. Within QCD sum rules, it is suggested that these heavy 
charmonium states interact with the nuclear medium through the gluon 
condensates. This is contrary to the interaction of the light vector mesons
($\rho$, $\omega$, $\phi$), which interact with the nuclear 
medium through the quark condensates. 
This is because all the heavy quark condensates can be related to the 
gluon condensates via heavy-quark expansion \cite{kimlee}. Also in the
nuclear medium there are no valence charm quarks to leading order in 
density and any interaction with the medium is gluonic. The QCD sum 
rules has been used to study the medium modification of $D$ mesons 
\cite{haya1} and light vector mesons \cite{hatsuda}. The QCD sum rule 
approach \cite{klingl} and leading order perturbative calculations 
\cite{pes1} to study the medium modifications of charmonium,
show that the mass of $J/\psi$ is reduced only slightly in the nuclear medium. 
In \cite{lee1}, the mass modification of charmonium has been studied using 
leading order QCD formula and the linear density approximation for
the gluon condensate in the nuclear medium. This shows a small drop
for the $J/\psi$ mass at the nuclear matter density, but there is seen
to be significant shift in the masses of the excited states of charmonium 
($\psi$(3686) and $\psi$(3770)). 

In the present work, we study the medium modification of the masses 
of $J/\psi$ and excited charmonium states $\psi(3686)$ and $\psi(3770)$ 
in the nuclear medium due to the interaction with the gluon condensates
using the leading order QCD formula. The gluon condensate in the nuclear 
medium is calculated from the medium modification of a scalar 
dilaton field introduced within a chiral SU(3) model \cite{papa}
through a scale symmetry breaking term in the Lagrangian density
leading to the QCD trace anomaly. In the chiral SU(3) model, the gluon 
condensate is related to the fourth power of the dilaton field $\chi$ 
and the changes in the dilaton field with density are seen to be small. 
%
The model has been used successfully to study the medium modifications 
of kaons and antikaons in isospin asymmetric nuclear matter in 
\cite{isoamss} and in hyperonic matter in \cite{isoamss2}. The model 
has also been used to study the $D$ mesons in asymmetric nuclear matter 
at zero temperature \cite{amarind} and in the symmetric and asymmetric 
nuclear matter at finite temperatures in Ref.\cite{amdmeson} and 
\cite{amarvind}. The vector mesons have also been studied within 
the model \cite{hartree, kristof1}. In the present investigation,
we study the isospin dependence of the in-medium masses
of charmonium obtained from the dilaton field, $\chi$
calculated for the asymmetric nuclear matter at finite temperatures.
This study will be of relevance for the experimental observables
from high density matter produced in the asymmetric nuclear collisions
at the future facility at GSI.

The outline of the paper is as follows : In section II, we give a brief 
introduction of the chiral $SU(3)$ model used to study the in-medium
masses of charmonium in the present investigation. The medium 
modifications of the charmonium masses arise from the medium 
modification of a scalar dilaton field introduced in the hadronic 
model to incorporate broken scale invariance of QCD leading to QCD 
trace anomaly. In section III, we summarize the results obtained 
in the present investigation.

\section{The hadronic chiral $SU(3) \times SU(3)$ model }
We use an effective chiral $SU(3)$ model for the present investigation 
\cite{papa}. The model is based on the nonlinear realization of chiral 
symmetry \cite{weinberg, coleman, bardeen} and broken scale invariance 
\cite{papa,hartree, kristof1}. This model has been used successfully to 
describe nuclear matter, finite nuclei, hypernuclei and neutron stars. 
The effective hadronic chiral Lagrangian density contains the following terms
\begin{equation}
{\cal L} = {\cal L}_{kin}+\sum_{W=X,Y,V,A,u} {\cal L}_{BW} + 
{\cal L}_{vec} + {\cal L}_{0} + {\cal L}_{SB}
\label{genlag}
\end{equation}
In Eq. (\ref{genlag}), ${\cal L}_{kin}$ is kinetic energy term, 
${\cal L}_{BW}$ is the baryon-meson interaction term in which the 
baryon-spin-0 meson interaction term generates the vacuum baryon masses. 
${\cal L}_{vec}$  describes the dynamical mass generation of the vector 
mesons via couplings to the scalar mesons and contain additionally 
quartic self-interactions of the vector fields. ${\cal L}_{0}$ contains 
the meson-meson interaction terms inducing the spontaneous breaking of 
chiral symmerty as well as a scale invariance breaking logarthimic 
potential. ${\cal L}_{SB}$ describes the explicit chiral symmetry breaking. 

To study the hadron properties at finite temperature and densities
in the present investigation, we use the mean  field approximation,
where all the meson fields are treated as classical fields. 
In this approximation, only the scalar and the vector fields 
contribute to the baryon-meson interaction, ${\cal L}_{BW}$
since for all the other mesons, the expectation values are zero.
The interactions of the scalar mesons and vector mesons with the
baryons are given as
\begin{eqnarray}
{\cal  L} _{Bscal} +  {\cal L} _{Bvec} = - \sum_{i} \bar{\psi}_{i} 
\left[  m_{i}^{*} + g_{\omega i} \gamma_{0} \omega 
+ g_{\rho i} \gamma_{0} \rho + g_{\phi i} \gamma_{0} \phi 
\right] \psi_{i}. 
\end{eqnarray}
The interaction of the vector mesons, of the scalar fields and 
the interaction corresponding to the explicitly symmetry breaking
in the mean field approximation are given as
\begin{eqnarray}
 {\cal L} _{vec} & = & \frac{1}{2} \left( m_{\omega}^{2} \omega^{2} 
+ m_{\rho}^{2} \rho^{2} + m_{\phi}^{2} \phi^{2} \right) 
\frac{\chi^{2}}{\chi_{0}^{2}}
\nonumber \\
& + &  g_4 (\omega ^4 +6\omega^2 \rho^2+\rho^4 + 2\phi^4),
\end{eqnarray}
\begin{eqnarray}
{\cal L} _{0} & = & -\frac{1}{2} k_{0}\chi^{2} \left( \sigma^{2} + \zeta^{2} 
+ \delta^{2} \right) + k_{1} \left( \sigma^{2} + \zeta^{2} + \delta^{2} 
\right)^{2} \nonumber\\
&+& k_{2} \left( \frac{\sigma^{4}}{2} + \frac{\delta^{4}}{2} + 3 \sigma^{2} 
\delta^{2} + \zeta^{4} \right) 
+ k_{3}\chi\left( \sigma^{2} - \delta^{2} \right)\zeta \nonumber\\
&-& k_{4} \chi^{4} - \frac{1}{4} \chi^{4} {\rm {ln}} 
\frac{\chi^{4}}{\chi_{0}^{4}} 
+ \frac{d}{3} \chi^{4} {\rm {ln}} \Bigg (\bigg( \frac{\left( \sigma^{2} 
- \delta^{2}\right) \zeta }{\sigma_{0}^{2} \zeta_{0}} \bigg) 
\bigg (\frac{\chi}{\chi_0}\bigg)^3 \Bigg ),
\label{lagscal}
\end{eqnarray}
and 
\begin{eqnarray}
{\cal L} _{SB} & = & - \left( \frac{\chi}{\chi_{0}}\right) ^{2} 
\left[ m_{\pi}^{2} 
f_{\pi} \sigma + \left( \sqrt{2} m_{k}^{2}f_{k} - \frac{1}{\sqrt{2}} 
m_{\pi}^{2} f_{\pi} \right) \zeta \right]. 
\end{eqnarray}
The effective mass of the baryon of species $i$ is given as
\begin{equation}
{m_i}^{*} = -(g_{\sigma i}\sigma + g_{\zeta i}\zeta + g_{\delta i}\delta)
\label{mbeff}
\end{equation}
The baryon-scalar meson interactions, as can be seen from equation
(\ref{mbeff}), generate the baryon masses through 
the coupling of  baryons to the non-strange $\sigma$, strange $\zeta$ 
scalar mesons and also to scalar-isovector meson $\delta$. In analogy 
to the baryon-scalar meson coupling there exist two independent 
baryon-vector meson interaction terms corresponding to the F-type 
(antisymmetric) and D-type (symmetric) couplings. Here antisymmetric 
coupling is used because the universality principle \cite{saku69} 
and vector meson dominance model suggest small symmetric coupling. 
Additionally,  we choose the parameters \cite{papa,isoamss} so as 
to decouple the strange vector field $\phi_{\mu}\sim\bar{s}\gamma_{\mu}s$ 
from the nucleon, corresponding to an ideal mixing between $\omega$ and 
$\phi$ mesons. A small deviation of the mixing angle from ideal mixing 
\cite{dumbrajs,rijken,hohler1} has not been taken into account in the 
present investigation.

The concept of broken scale invariance leading to the trace anomaly 
in (massless) QCD, $\theta_{\mu}^{\mu} = \frac{\beta_{QCD}}{2g} 
{G^a}_{\mu\nu} G^{\mu\nu a}$, where $G_{\mu\nu}^{a} $ is the 
gluon field strength tensor of QCD, is simulated in the effective 
Lagrangian at tree level \cite{sche1} through the introduction of 
the scale breaking terms 
\begin{equation}
{\cal L}_{scalebreaking} =  -\frac{1}{4} \chi^{4} {\rm {ln}}
\Bigg ( \frac{\chi^{4}} {\chi_{0}^{4}} \Bigg ) + \frac{d}{3}{\chi ^4} 
{\rm {ln}} \Bigg ( \bigg (\frac{I_{3}}{{\rm {det}}\langle X 
\rangle _0} \bigg ) \bigg ( \frac {\chi}{\chi_0}\bigg)^3 \Bigg ),
\label{scalebreak}
\end{equation}
where $I_3={\rm {det}}\langle X \rangle$, with $X$ as the multiplet
for the scalar mesons. These scale breaking terms,
in the mean field approximation, are given by the last two terms
of the Lagrangian density, ${\cal L}_0$  given by equation (\ref{lagscal}). 
The effect of these logarithmic terms is to break the scale invariance, 
which leads to the trace of the energy momentum tensor as \cite{heide1}
\begin{equation}
\theta_{\mu}^{\mu} = \chi \frac{\partial {\cal L}}{\partial \chi} 
- 4{\cal L} 
= -(1-d)\chi^{4}.
\label{tensor1}
\end{equation}
Hence the scalar gluon condensate of QCD ($\langle {G^a}_{\mu \nu}
G^{\mu \nu a} \rangle$) is simulated by a scalar dilaton field in the present
hadronic model. 

The coupled equations of motion for the non-strange scalar field $\sigma$, 
strange scalar field $ \zeta$, scalar-isovector field $ \delta$ and dilaton 
field $\chi$, are derived from the Lagrangian density
and are given as
\begin{eqnarray}
&& k_{0}\chi^{2}\sigma-4k_{1}\left( \sigma^{2}+\zeta^{2}
+\delta^{2}\right)\sigma-2k_{2}\left( \sigma^{3}+3\sigma\delta^{2}\right)
-2k_{3}\chi\sigma\zeta \nonumber\\
&-&\frac{d}{3} \chi^{4} \bigg (\frac{2\sigma}{\sigma^{2}-\delta^{2}}\bigg )
+\left( \frac{\chi}{\chi_{0}}\right) ^{2}m_{\pi}^{2}f_{\pi}
-\sum g_{\sigma i}\rho_{i}^{s} = 0 
\label{sigma}
\end{eqnarray}
\begin{eqnarray}
&& k_{0}\chi^{2}\zeta-4k_{1}\left( \sigma^{2}+\zeta^{2}+\delta^{2}\right)
\zeta-4k_{2}\zeta^{3}-k_{3}\chi\left( \sigma^{2}-\delta^{2}\right)\nonumber\\
&-&\frac{d}{3}\frac{\chi^{4}}{\zeta}+\left(\frac{\chi}{\chi_{0}} \right) 
^{2}\left[ \sqrt{2}m_{k}^{2}f_{k}-\frac{1}{\sqrt{2}} m_{\pi}^{2}f_{\pi}\right]
 -\sum g_{\zeta i}\rho_{i}^{s} = 0 
\label{zeta}
\end{eqnarray}
\begin{eqnarray}
& & k_{0}\chi^{2}\delta-4k_{1}\left( \sigma^{2}+\zeta^{2}+\delta^{2}\right)
\delta-2k_{2}\left( \delta^{3}+3\sigma^{2}\delta\right) +k_{3}\chi\delta 
\zeta \nonumber\\
& + &  \frac{2}{3} d \left( \frac{\delta}{\sigma^{2}-\delta^{2}}\right)
-\sum g_{\delta i}\rho_{i}^{s} = 0
\label{delta}
\end{eqnarray}
 
\begin{eqnarray}
& & k_{0}\chi \left( \sigma^{2}+\zeta^{2}+\delta^{2}\right)-k_{3}
\left( \sigma^{2}-\delta^{2}\right)\zeta + \chi^{3}\left[1
+{\rm {ln}}\left( \frac{\chi^{4}}{\chi_{0}^{4}}\right)  \right]
+(4k_{4}-d)\chi^{3}
\nonumber\\
& - & \frac{4}{3} d \chi^{3} {\rm {ln}} \Bigg ( \bigg (\frac{\left( \sigma^{2}
-\delta^{2}\right) \zeta}{\sigma_{0}^{2}\zeta_{0}} \bigg ) 
\bigg (\frac{\chi}{\chi_0}\bigg)^3 \Bigg ) 
+\frac{2\chi}{\chi_{0}^{2}}\left[ m_{\pi}^{2}
f_{\pi}\sigma +\left(\sqrt{2}m_{k}^{2}f_{k}-\frac{1}{\sqrt{2}}
m_{\pi}^{2}f_{\pi} \right) \zeta\right]  = 0 
\label{chi}
\end{eqnarray}
In the above, ${\rho_i}^s$ are the scalar densities for the baryons, 
given as 
\begin{eqnarray}
\rho_{i}^{s} = \gamma_{i}\int\frac{d^{3}k}{(2\pi)^{3}} 
\frac{m_{i}^{*}}{E_{i}^{*}(k)} 
\Bigg ( \frac {1}{e^{({E_i}^* (k) -{\mu_i}^*)/T}+1}
+ \frac {1}{e^{({E_i}^* (k) +{\mu_i}^*)/T}+1} \Bigg )
\label{scaldens}
\end{eqnarray}
where, ${E_i}^*(k)=(k^2+{{m_i}^*}^2)^{1/2}$, and, ${\mu _i}^* 
=\mu_i -g_{\omega i}\omega -g_{\rho i}\rho -g_{\phi i}\phi$, are the single 
particle energy and the effective chemical potential
for the baryon of species $i$, and,
$\gamma_i$=2 is the spin degeneracy factor \cite{isoamss}.

The above coupled equations of motion are solved to obtain the density 
and temperature dependent values of the scalar fields ($\sigma$,
$\zeta$ and $\delta$) and the dilaton field, $\chi$, in the isospin
asymmetric hot nuclear medium. As has been already mentioned, the value 
of the $\chi$ is related to the scalar gluon condensate in the hot 
hadronic medium, and is used to compute the in-medium masses of charmonium 
states, in the present investigation. The isospin asymmetry in the medium
is introduced through the scalar-isovector field $\delta$ 
and therefore the dilaton field obtained after solving the above 
equations is also dependent on the isospin asymmetry parameter,
$\eta$ defined as $\eta= ({\rho_n -\rho_p})/({2 \rho_B})$, 
where $\rho_n$ and $\rho_p$ are the number densities of the neutron
and the proton and $\rho_B$ is the baryon density. In the present 
investigation, we study the effect of isospin asymmetry of the medium 
on the masses of the charmonium states $J/\psi, \psi(3686)$ 
and $\psi(3770)$.

The comparison of the trace of the energy momentum tensor arising
from the trace anomaly of QCD with that of the present chiral model
gives the relation of the dilaton field to the scalar gluon condensate.
We have, in the limit of massless quarks \cite{cohen},
\begin{equation}
\theta_{\mu}^{\mu} = \langle \frac{\beta_{QCD}}{2g} 
G_{\mu\nu}^{a} G^{\mu\nu a} \rangle  \equiv  -(1 - d)\chi^{4} 
\label{tensor2}
\end{equation}
The parameter $d$ originates from the second logarithmic term of equation 
(\ref{scalebreak}). To get an insight into the value of the parameter 
$d$, we recall that the QCD $\beta$ function at one loop level, for 
$N_{c}$ colors and $N_{f}$ flavors is given by
\begin{equation}
\beta_{\rm {QCD}} \left( g \right) = -\frac{11 N_{c} g^{3}}{48 \pi^{2}} 
\left( 1 - \frac{2 N_{f}}{11 N_{c}} \right)  +  O(g^{5})
\label{beta}
\end{equation}
In the above equation, the first term in the parentheses arises from 
the (antiscreening) self-interaction of the gluons and the second term, 
proportional to $N_{f}$, arises from the (screening) contribution of 
quark pairs. Equations (\ref{tensor2}) and (\ref{beta}) suggest the 
value of $d$ to be 6/33 for three flavors and three colors, and 
for the case of three colors and two flavors, the value of $d$ 
turns out to be 4/33, to be consistent with the one loop estimate 
of QCD $\beta$ function. These values give the order of magnitude 
about which the parameter $d$ can be taken \cite{heide1}, since one 
cannot rely on the one-loop estimate for $\beta_{\rm {QCD}}(g)$. 
In the present investigation of the in-medium properties of the 
charmonium states due to the medium modification of the dilaton 
field within chiral $SU(3)$ model, we use the value of $d$=0.064
\cite{amarind}. This parameter, along with the other parameters
corresponding to the  scalar Lagrangian density, ${\cal L}_0$ 
given by ({\ref{lagscal}), are fitted so as to ensure 
extrema in the vacuum for the $\sigma$, $\zeta$ and $\chi$ field 
equations, to  reproduce the vacuum masses of the $\eta$ and $\eta '$ 
mesons, the mass of the $\sigma$ meson around 500 MeV, and,
pressure, p($\rho_0$)=0,
with $\rho_0$ as the nuclear matter saturation density \cite{papa,amarind}.

The trace of the energy-momentum tensor in QCD, using the 
one loop beta function given by equation (\ref{beta}),
for $N_c$=3 and $N_f$=3, is given as,
\begin{equation}
\theta_{\mu}^{\mu} = - \frac{9}{8} \frac{\alpha_{s}}{\pi} 
G_{\mu\nu}^{a} G^{\mu\nu a}
\label{tensor4}
\end{equation} 
Using equations (\ref{tensor2}) and (\ref{tensor4}), we can write  
\begin{equation}
\left\langle  \frac{\alpha_{s}}{\pi} G_{\mu\nu}^{a} G^{ \mu\nu a} 
\right\rangle =  \frac{8}{9}(1 - d) \chi^{4}
\label{chiglu}
\end{equation}
We thus see from the equation (\ref{chiglu}) that the scalar 
gluon condensate $\left\langle \frac{\alpha_{s}}{\pi} G_{\mu\nu}^{a} 
G^{\mu\nu a}\right\rangle$ is proportional to the fourth power of the 
dilaton field, $\chi$, in the chiral SU(3) model.
As mentioned earlier, the in-medium masses of charmonium states are 
modified due to the gluon condensates. Therefore, we need to know the 
change in the gluon condensate with density and temperature
of the asymmetric nuclear medium, which is calculated from the 
modification of the $\chi$ field, by using equation (\ref{chiglu}). 

From the QCD sum rule calculations, the mass shift of the charmonium 
states in the medium is due to the gluon condensates \cite{haya1,lee1}.
For heavy quark systems, there are two independent lowest dimension
operators: the scalar gluon condensate ( $\left\langle 
\frac{\alpha_{s}}{\pi} G_{\mu\nu}^{a} G^{\mu\nu a}\right\rangle$ ) 
and the condensate of the twist 2 gluon operator ( $\left\langle 
\frac{\alpha_{s}}{\pi} G_{\mu\nu}^{a} G^{\mu\alpha a}\right\rangle$ ). 
These operators can be rewritten in terms of the color electric and 
color magnetic fields, $\langle \frac{\alpha_s}{\pi} {\vec E}^2\rangle$ 
and $\langle \frac{\alpha_s}{\pi} {\vec B}^2\rangle$. Additionally, 
since the Wilson coefficients for the operator $\langle \frac{\alpha_s}{\pi} 
{\vec B}^2\rangle$ vanishes in the non-relativistic limit, the only 
contribution from the gluon condensates is proportional to $\langle 
\frac{\alpha_s}{\pi} {\vec E}^2\rangle$, similar to the second order 
Stark effect.  Hence, the mass shift of the charmonium states  arises
due to the change in the operator $\langle \frac{\alpha_s}{\pi} 
{\vec E}^2\rangle$ in the medium from its vacuum value \cite {lee1}. 
In the leading order mass shift formula derived in the large charm 
mass limit \cite{pes1}, the shift in the mass of the charmonium 
state is given as
\cite{lee1}
\begin{equation}
\Delta m_{\psi} (\epsilon) = -\frac{1}{9} \int dk^{2} \vert 
\frac{\partial \psi (k)}{\partial k} \vert^{2} \frac{k}{k^{2} 
/ m_{c} + \epsilon} \bigg ( 
\left\langle  \frac{\alpha_{s}}{\pi} E^{2} \right\rangle-
\left\langle  \frac{\alpha_{s}}{\pi} E^{2} \right\rangle_{0}
\bigg ).
\label{mass1}
\end{equation}
In the above, $m_c$ is the mass of the charm quark, taken as 1.95 GeV 
\cite{lee1}, $m_\psi$ is the vacuum mass of the charmonium state 
and $\epsilon = 2 m_{c} - m_{\psi}$. 
$\psi (k)$ is the wave function of the charmonium state
in the momentum space, normalized as $\int\frac{d^{3}k}{2\pi^{3}} 
\vert \psi(k) \vert^{2} = 1 $ \cite{leetemp}.
At finite densities, in the linear density approximation, the change 
in the value of $\langle \frac{\alpha_s}{\pi} {\vec E}^2\rangle$, 
from its vacuum value, is given as 
\begin{equation}
\left\langle  \frac{\alpha_{s}}{\pi} E^{2} \right\rangle-
\left\langle  \frac{\alpha_{s}}{\pi} E^{2} \right\rangle_{0}
=
\left\langle  \frac{\alpha_{s}}{\pi} E^{2} \right\rangle _{N}
\frac {\rho_B}{2 M_N},
\end{equation}
and the mass shift in the charmonium states reduces to \cite{lee1}
\begin{equation}
\Delta m_{\psi} (\epsilon) = -\frac{1}{9} \int dk^{2} \vert 
\frac{\partial \psi (k)}{\partial k} \vert^{2} \frac{k}{k^{2} 
/ m_{c} + \epsilon} 
\left\langle  \frac{\alpha_{s}}{\pi} E^{2} \right\rangle _{N}
\frac {\rho_B}{2 M_N}.
\label{masslindens}
\end{equation}
In the above, $\left\langle  \frac{\alpha_{s}}{\pi} E^{2} 
\right\rangle _{N}$ is the expectation value of  
$\left\langle  \frac{\alpha_{s}}{\pi} E^{2} \right\rangle$
with respect to the nucleon.

The expectation value of the scalar gluon condensate can be expressed 
in terms of the color electric field and the color magnetic field 
as \cite{david}
\begin{equation}
\left\langle 
\frac{\alpha_{s}}{\pi} G_{\mu\nu}^{a} G^{\mu\nu a}\right\rangle 
=-2 \left\langle \frac{\alpha_{s}}{\pi} (E^{2} - B^{2}) \right\rangle.
\end{equation}
In the non-relativistic limit, as already mentioned, the contribution
from the magnetic field vanishes and hence, we can write,
\begin{equation}
\left\langle \frac{\alpha_{s}}{\pi} E^{2} \right\rangle
=-\frac {1}{2} 
\left\langle \frac{\alpha_{s}}{\pi} 
G_{\mu\nu}^{a} G^{\mu\nu a}\right\rangle 
\label{e2glu}
\end{equation}

Using equations (\ref{chiglu}), (\ref{mass1}) and (\ref{e2glu}), 
we obtain the expression for the mass shift in the charmonium 
in the hot and dense nuclear medium, which arises from 
the change in the dilaton field in the present investigation, 
as
\begin{equation}
\Delta m_{\psi} (\epsilon) = \frac{4}{81} (1 - d) \int dk^{2} 
\vert \frac{\partial \psi (k)}{\partial k} \vert^{2} \frac{k}{k^{2} 
/ m_{c} + \epsilon}  \left( \chi^{4} - {\chi_0}^{4}\right). 
\label{masspsi}
\end{equation}
In the above, $\chi$ and $\chi_0$ are the values of the dilaton field
in the nuclear medium and the vacuum respectively.

In the present investigation, the wave functions for the charmonium states 
are taken to be Gaussian and are given as \cite{charmwavefn}
\begin{equation}
\psi_{N, l} = Normalization \times Y_{l}^{m} (\theta, \phi) 
(\beta^{2} r^{2})^{\frac{1}2{} l} exp^{-\frac{1}{2} \beta^{2} r^{2}} 
L_{N - 1}^{l + \frac{1}{2}} \left( \beta^{2} r^{2}\right)
\label{wavefn} 
\end{equation} 
where $\beta^{2} = M \omega / h$ characterizes the strength of the 
harmonic potential, $M = m_{c}/2$ is the reduced mass of 
the charm quark and charm anti-quark system, and $L_{p}^{k} (z)$ 
is the associated Laguerre Polynomial. As in Ref. \cite{lee1},
the oscillator constant $\beta$ is determined from the mean squared 
radii $\langle r^{2} \rangle$ as 0.46$^{2}$ fm$^2$, 0.96$^{2}$ fm$^2$ 
and 1 fm$^{2}$ for the charmonium states $J/\psi(3097) $, $\psi(3686)$ and 
$\psi(3770)$, respectively. This gives the value for the parameter
$\beta$ as 0.51 GeV, 0.38 GeV and 0.37 GeV for $J/\psi(3097)$, 
$\psi(3686$ and $\psi(3770)$, assuming that these 
charmonium states are in the 1S, 2S and 1D states respectively. 
Knowing the wave functions of the charmonium states and 
calculating the medium modification of the dilaton field
in the hot nuclear matter, we obtain the mass shift of the
charmonium states, $J/\psi$, $\psi (3686)$ and $\psi (3770)$
respectively. In the next section we shall present the results 
of the present investigation of these in-medium charmonium masses 
in hot asymmetric nuclear matter.

\section{Results and Discussions}

In this section, we first investigate the effects of density, 
isospin-asymmetry and temperature of the nuclear medium on the dilaton 
field $\chi$ using a chiral SU(3) model. From the medium modification 
of the $\chi$ field, we shall then study the in-medium masses of 
charmonium states $J/\psi$, $\psi(3686)$ and $\psi(3770)$ using 
equation (\ref{masspsi}). 

The values of the parameters used in the present investigation, 
are : $k_{0} = 2.54, k_{1} = 1.35, k_{2} = -4.78, k_{3} = -2.77$, 
$k_{4} = -0.22$ and $d =  0.064$, which are the parameters
occurring in the scalar meson interactions defined in equation 
(\ref{lagscal}). 
The vacuum values of the scalar isoscalar fields, $\sigma$ and $\zeta$ 
and the dilaton field $\chi$ are $-93.3$ MeV, $-106.6$ MeV and 409.77 MeV
respectively. 
The values, $g_{\sigma N} = 10.6$ and $g_{\zeta N} = -0.47$ are 
determined by fitting to the vacuum baryon masses. The other parameters 
fitted to the asymmetric nuclear matter saturation properties 
in the mean-field approximation are: $g_{\omega N}$ = 13.3, 
$g_{\rho p}$ = 5.5, $g_{4}$ = 79.7, $g_{\delta p}$ = 2.5, 
$m_{\zeta}$ = 1024.5 MeV, $ m_{\sigma}$ = 466.5 MeV 
and $m_{\delta}$ = 899.5 MeV. The nuclear matter saturation 
density used in the present investigation is $0.15$ fm$^{-3}$.
 
\begin{figure}
\includegraphics[width=16cm,height=16cm]{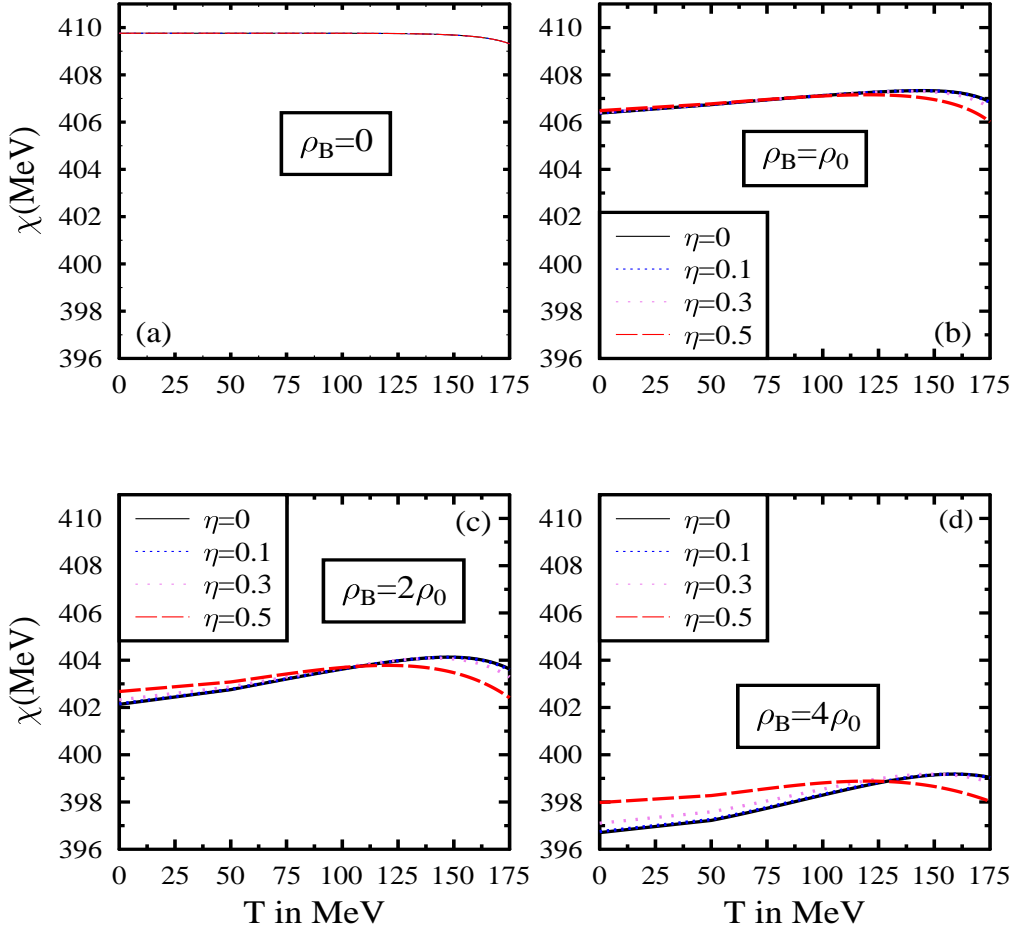}
\caption{(Color online)
The dilaton field $\chi$ plotted as a function of the
temperature, at given baryon densities,
for different values of the isospin asymmetry parameter, $\eta$.
}
\label{chitemp}
\end{figure}

The variations of the dilaton field $\chi$ with density, temperature
and isospin asymmetry, within the chiral SU(3) model, are obtained by 
solving the coupled equations of motion of scalar fields given by 
equations (\ref{sigma}), (\ref{zeta}), (\ref{delta}) and (\ref{chi}). 
In figure \ref{chitemp}, we show the variation of dilaton field $\chi$, 
with temperature, for both zero and finite baryon densities,
and for selected values of the isospin asymmetry parameter, 
$\eta$ = 0, 0.1, 0.3 and 0.5. At zero baryon density, it is observed 
that the value of the dilaton field remains almost 
a constant upto a temperature of about 130 MeV above which it is 
seen to drop with increase in temperature. However, the drop 
in the dilaton field is seen to be very small. The value of the 
dilaton field is seen to change from 409.8 MeV at T=0 to about 
409.7 MeV and 409.3 MeV at T=150 MeV and T=175 MeV respectively.
The thermal distribution functions have an effect of increasing 
the scalar densities at zero baryon density, i.e., for $\mu_i ^*$=0, 
as can be seen from the expression of the scalar densities, given 
by (\ref{scaldens}). This effect seems to be negligible upto
a temperature of about 130 MeV. This leads to a decrease in 
the magnitudes of scalar fields, $\sigma$ and $\zeta$. 
This behaviour of the scalar fields is reflected in the value of $\chi$, 
which is solved from the coupled equations of motion of the 
scalar fields, given by equations (\ref{sigma}), (\ref{zeta}), 
(\ref{delta}) and (\ref{chi}), as a drop as we 
increase the temperature above a temperature of about 130 MeV.
The scalar densities attaining nonzero values at high temperatures, 
even at zero baryon density, indicates the presence of baryon-antibaryon 
pairs in the thermal bath and has already been observed in the literature
\cite{kristof1,scalartemp}. This leads 
to the baryon masses to be different from their vacuum masses above this 
temperature, arising from modifications of the scalar fields $\sigma$ and
$\zeta$.

For finite density situations, the behaviour of the $\chi$ field
with temperature is seen to be very different from the zero density
case, as can be seen from the subplots (b),(c) and (d) of figure \ref{chitemp}, 
where the $\chi$ field is plotted as a function of the temperature
for densities $\rho_0$, 2$\rho_0$ and 4$\rho_0$ respectively. 
At finite densities, one observes first a rise and then a decrease of 
the dilaton field with temperature. This is related to the fact that 
at finite densities, the magnitude of the $\sigma$ field (as well as 
of the $\zeta$ field) first show an increase and then a drop with further 
increase of the temperature \cite{amarvind} which is reflected in the
behaviour of $\chi$ field, since it is solved from the coupled 
equations of the scalar fields. The reason for the different behaviour 
of the scalar fields ($\sigma$ and $\zeta$) at zero and finite densities
can be understood in the following manner \cite{kristof1}. As has been
already mentioned, the thermal distribution functions in (\ref{scaldens}) 
have an effect of increasing the scalar densities at zero baryon density, i.e., 
for $\mu_i ^*$=0. However, at finite densities, i.e., for nonzero values 
of the effective chemical potential, ${\mu_i}^*$, for increasing temperature, 
there are contributions also from higher momenta, thereby, 
increasing the denominator of the integrand on the right hand side of
the equation (\ref{scaldens}). This leads to a decrease in the scalar 
density. The competing effects of the thermal distribution functions
and the contributions of the higher momenta states
give rise to the observed effect of the scalar density
and hence of the $\sigma$ and $\zeta$ fields with temperature
at finite baryon densities \cite{kristof1}. This kind of behaviour 
of the scalar $\sigma$ field on temperature at finite densities 
has also been observed in the Walecka model by Li and Ko \cite{liko}, 
which was reflected as an increase in the mass of the nucleon with 
temperature at finite densities in the mean field calculations.
The effects of the behaviour of the scalar fields on the value 
of the $\chi$ field, obtained from solving the coupled equations 
(\ref{sigma}) to (\ref{chi}) for the scalar fields, are shown in 
figure \ref{chitemp}. 

In figure \ref{chitemp}, it is observed that for a given value of 
isospin asymmetry parameter $\eta$, the dilaton field $\chi$ decreases 
with increase in the density of the nuclear medium. The drop 
in the value of $\chi$ with density is seen to be much larger as compared
to its modification with temperature at a given density. 
For isospin symmetric nuclear medium ($\eta = 0$) at temperature $T = 0$, 
the reduction in the dilaton field $\chi$ from its vacuum value ($\chi_0$
 = 409.8 MeV), is seen to be about 3 MeV at $\rho_{B} = \rho_{0}$ 
and about 13 MeV, for $\rho_{B} = 4\rho_{0}$. 
As we move from isospin symmetric medium, with $\eta = 0$, to isospin 
asymmetric medium, at temperature $T = 0$, and, for a given value 
of density, there is seen to be an increase in the value of the 
dilaton field $\chi$. However, the effect of isospin asymmetry 
of the medium on the value of the dilaton field is observed to be
negligible upto about a density of nuclear matter saturation density, 
and is appreciable only at higher values of densities as can be seen 
in figure \ref{chitemp}. At nuclear matter saturation density, 
$\rho_{0}$, the value of dilaton field $\chi$ changes from 
$406.4$ MeV in symmetric nuclear medium ($\eta = 0$) to $406.5$ MeV in 
the isospin asymmetric nuclear medium ($\eta = 0.5$). At a density of 
about $4\rho_{0}$, the values of the dilaton field are modified to 
396.7 MeV and 398 MeV at $\eta =0$ and $0.5$, respectively. Thus the 
increase in the dilaton field $\chi$ with isospin asymmetry of the 
medium is seen to be more at zero temperature as we move to 
higher densities.


At a finite density, $\rho_{B}$, and for given isospin asymmetry 
parameter $\eta$, the dilaton field $\chi$ is seen to first increase 
with temperature and above a particular value of the temperature, 
it is seen to decrease with further increase in temperature. At the nuclear 
saturation density $\rho_{B} = \rho_{0}$ and in isospin symmetric 
nuclear medium ($\eta = 0$) the value of the dilaton field $\chi$ 
increases upto a temperature of about $T = 145$ MeV, above with there
is a drop in the dilaton field. For $\rho_B$=$\rho_0$ in the asymmetric 
nuclear matter with $\eta = 0.5$, there is seen to be a rise in the value 
of $\chi$ upto a temperature of about 120 MeV, above which it starts 
decreasing. As it has been already mentioned, at zero temperature and for a 
given value of density, the dilaton field $\chi$ is found to increase 
with increase in the isospin asymmetry of the nuclear medium. But 
from figure \ref{chitemp}, it is observed that at high temperatures 
and for a given density, the value of the dilaton field $\chi$ becomes 
higher in symmetric nuclear medium as compared to isospin asymmetric 
nuclear medium e.g. at nuclear saturation density $\rho_{B} = \rho_{0}$ 
and temperature $T = 150$ MeV the values of dilaton field $\chi$ are 
$407.3$ MeV and $407$ MeV at $\eta = 0$ and $0.5$ respectively. 
At density $\rho_{B} = 4 \rho_{0}$, $T = 150$ MeV the values of 
dilaton field $\chi$ are seen to be $399.1$ MeV and $398.7$ MeV 
for $\eta = 0$ and $0.5$ respectively. This observed behaviour 
of the $\chi$ is related
to the fact that at finite densities and for isospin asymmetric matter, 
there are contributions from the scalar isovector $\delta$ field,
whose magnitude is seen to decrease for higher temperatures
for given densities, whereas $\delta$ field has zero contribution
for isospin symmetric matter.

\begin{figure}
\includegraphics[width=16cm,height=16cm]{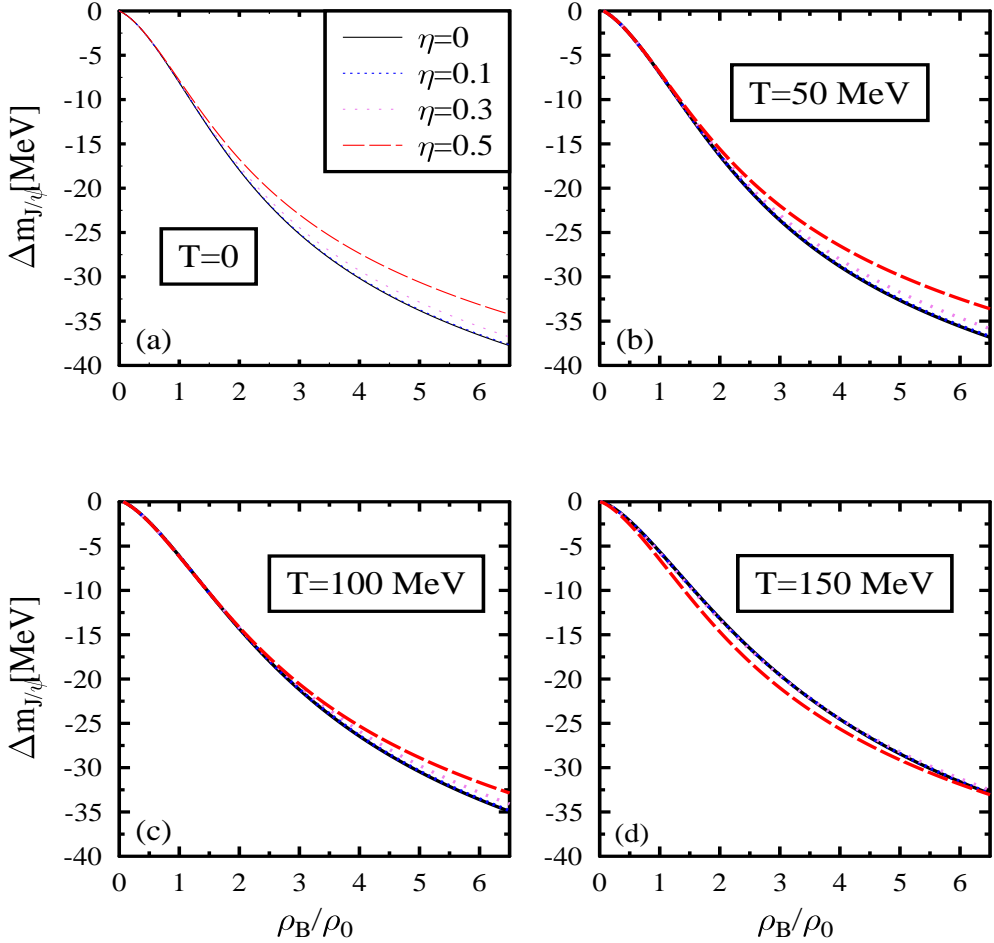}
\caption{(Color online)
The mass shift of J/$\psi$ plotted as a function of the
baryon density in units of nuclear matter saturation density
at given temperatures,
for different values of the isospin asymmetry parameter, $\eta$.
}
\label{mjpsi}
\end{figure}

\begin{figure}
\includegraphics[width=16cm,height=16cm]{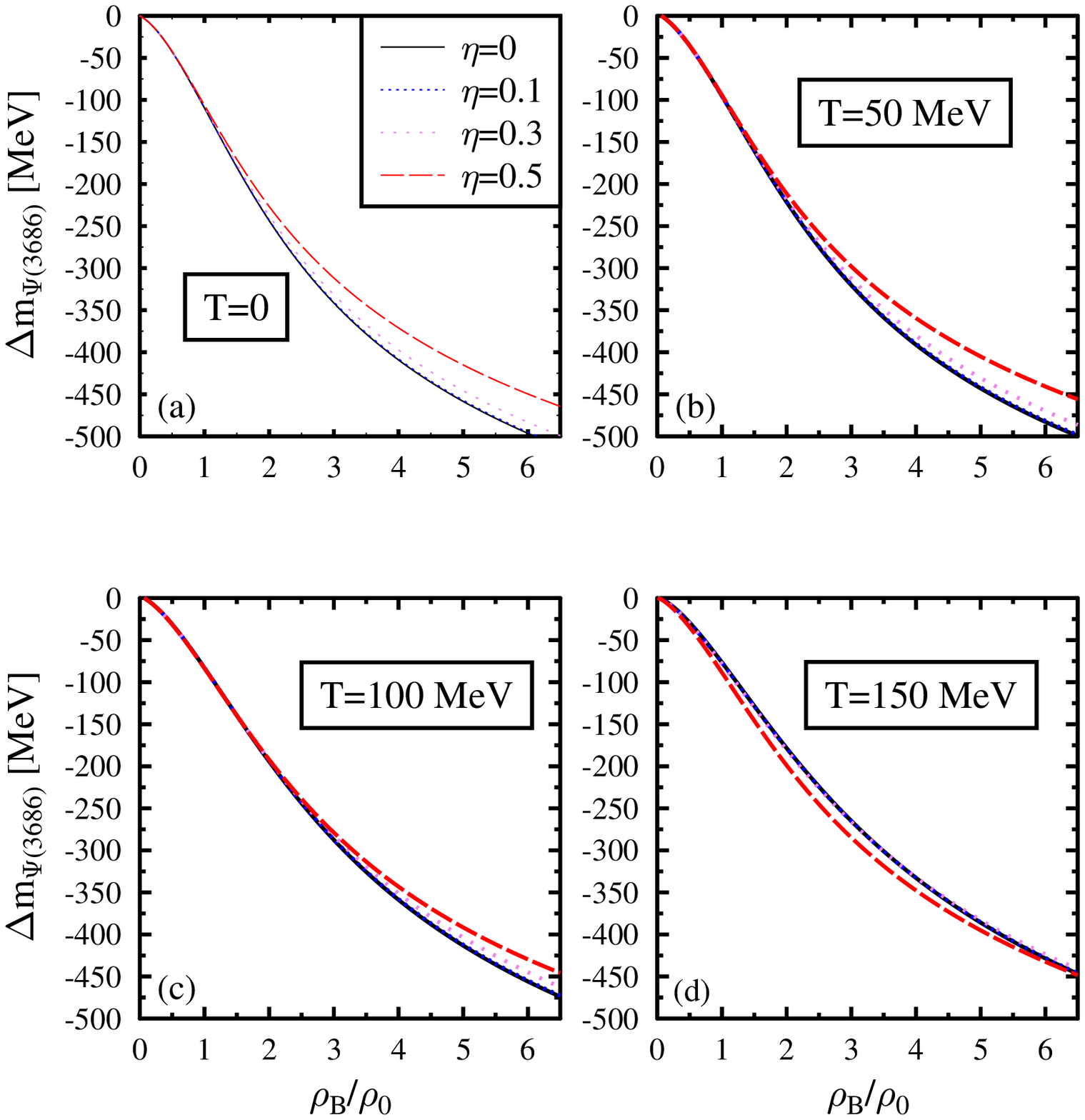}
\caption{(Color online)
The mass shift of $\psi$(3686) plotted as a function of the
baryon density in units of nuclear matter saturation density
at given temperatures,
for different values of the isospin asymmetry parameter, $\eta$.
}
\label{mpsi1}
\end{figure}

\begin{figure}
\includegraphics[width=16cm,height=16cm]{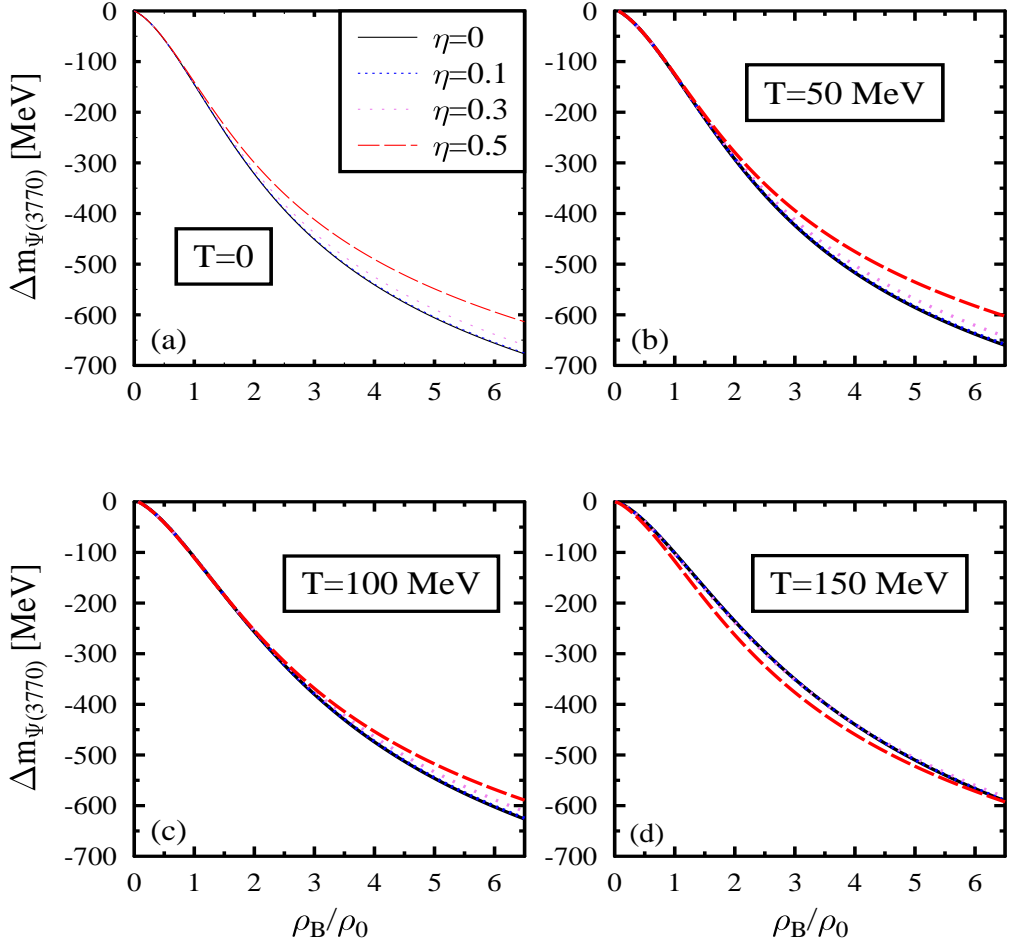}
\caption{(Color online)
The mass shift of $\psi$(3770) plotted as a function of the
baryon density in units of nuclear matter saturation density
at given temperatures,
for different values of the isospin asymmetry parameter, $\eta$.
}
\label{mpsi2}
\end{figure}

We shall now investigate how the behaviour of the dilaton field $\chi$ 
in the hot asymmetric nuclear medium affects the in-medium masses of 
the charmonium states $J/\psi$, $\psi(3686)$ and $\psi(3770)$. 
In figures \ref{mjpsi}, \ref{mpsi1} and \ref{mpsi2}, 
we show the shifts of the masses of charmonium states $J/\psi$, $\psi(3686)$ 
and $\psi(3770)$ from their vacuum values, as functions of the baryon density 
for given values of temperature $T$, and for different values of the
isospin asymmetry parameter, $\eta$. We have shown the results for the 
values of the temperature, T = 0, 50, 100 and 150 MeV. At the nuclear matter 
saturation density, $\rho_{B}$ = $\rho_{0}$ at temperature $T = 0$, 
the mass-shift for $J/\psi$ meson is observed to be about $-8.6$ MeV 
in the isospin symmetric nuclear medium ($\eta = 0$) 
and in the asymmetric nuclear medium, with isospin asymmetry parameter 
$\eta = 0.5$, it is seen to be $-8.4$ MeV. At $\rho_{B} = 4\rho_{0}$, 
temperature $T = 0$, the mass-shift  for $J/\psi$ meson is $-32.2$ MeV 
in the isospin symmetric nuclear medium ($\eta = 0$) and in isospin 
asymmetric nuclear medium ($\eta = 0.5$), it changes to $-29.2$ MeV. 
The increase in the magnitude of the mass-shift, with density $\rho_{B}$, 
is because of the larger drop in the dilaton field $\chi$ at higher densities. 
However, with increase in the isospin asymmetry of the medium the magnitude
of the mass-shift decreases because the drop in the dilaton field $\chi$ 
is less at a higher value of the isospin asymmetry parameter $\eta$.
At the nuclear matter saturation density $\rho_{B} = \rho_{0}$, 
and for temperature $T = 0$, the mass-shift for $\psi(3686)$ 
is observed to be about $-117$ and $-114$ MeV for values of the
$\eta = 0$ and $0.5$ respectively, and for $\psi(3770)$, 
the values of the mass-shift are seen to be about $-155$ MeV and 
$-$150 MeV respectively. 
At $\rho_{B} = 4\rho_{0}$ and zero temperature, the values 
of the mass-shift for $\psi(3686)$ are modified to $-436$ MeV 
and $-396$ MeV for $\eta = 0$ and $0.5$ respectively, and,
for $\psi(3770)$, the drop in the masses are about
 $-577$ MeV and $-523$ MeV respectively.
As mentioned earlier, the drop in the dilaton field, $\chi$, 
at finite temperature is less than at zero temperature and 
this behaviour is reflected in the smaller mass-shifts of the 
charmonium states at finite temperatures as compared to the zero 
temperature case. At nuclear matter saturation density $\rho_{B} = \rho_{0}$, 
and for temperature $T = 100$ MeV, the values of the the mass-shift for 
the $J/\psi$ meson are observed to be about $-6.77$ MeV and 
$-6.81$ MeV for isospin symmetric ($\eta = 0$) 
and isospin asymmetric ($\eta = 0.5$) nuclear medium respectively. 
At baryon density $\rho_{B} = 4\rho_{0}$, temperature $T = 100$ MeV, 
the mass-shift for $J/\psi$ is observed to be $-28.4$ MeV and $-27.2$ MeV 
for isospin symmetric ($\eta = 0$) and isospin asymmetric ($\eta = 0.5$) 
nuclear medium respectively. For the excited charmonium states $\psi(3686)$ 
and $\psi(3770)$, the mass-shifts at nuclear matter saturation density 
$\rho_{B} = \rho_{0}$ and temperature $T = 100$ MeV, are observed to
be $-91.8$ MeV and $-121.4$ MeV respectively, for isospin symmetric nuclear 
matter ($\eta = 0$) and $-92.4$ MeV and $-122$ MeV for the isospin 
asymmetric nuclear medium with $\eta = 0.5$. 
For baryon density $\rho_{B} = 4\rho_{0}$, 
and temperature $T = 100$ MeV, mass-shift for the charmonium states 
$\psi(3686)$ and $\psi(3770)$ are modified to about $-386$ MeV and $-510$
MeV respectively, for isospin symmetric ($\eta = 0$) and 
$-369$ MeV and $-488$ MeV for isospin asymmetric 
nuclear medium with $\eta = 0.5$. 

For temperature $T = 150$ MeV and at the nuclear matter saturation 
density $\rho_{B} = \rho_{0}$, the mass-shifts for the charmonium states 
$J/\psi, \psi(3686)$ and $\psi(3770)$ are seen to be $-6.25$, $-85$ 
and $-112$ MeV respectively in the isospin symmetric nuclear 
medium ($\eta = 0$). These values are modified to $-7.2$, $-98$ 
and $-129$ MeV respectively, in the isospin asymmetric nuclear medium 
with $\eta = 0.5$. At a baryon density $\rho_{B} = 4\rho_{0}$, 
the values of the mass-shift for $J/\psi, \psi(3686)$ and $\psi(3770)$ 
are observed to be $-26.4$, $-358$ and $-473$ MeV in isospin 
symmetric nuclear medium ($\eta = 0$) and in isospin asymmetric 
nuclear medium with $\eta = 0.5$, these values 
are modified to $-27.6$, $-375$ and $-494$ MeV respectively. 
Note that at high temperatures e.g at $T = 150$ MeV the mass-shift 
in isospin asymmetric nuclear medium ($\eta = 0.5$) is more as 
compared to isospin symmetric nuclear medium ($\eta = 0$). This is 
opposite to what is observed for the zero temperature case. 
The reason is that at high 
temperatures the dilaton field $\chi$ has more drop in the isospin 
asymmetric nuclear medium ($\eta = 0.5$) as compared to the isospin 
symmetric nuclear medium ($\eta = 0$), due to the contributions
from the $\delta$ field for nonzero $\eta$, which is observed to 
decrease in its magnitude at high temperatures.

The values of the mass-shift for the charmonium states obtained 
within the present investigation, at nuclear saturation density 
$\rho_{0}$ and temperature $T =0$, are in good agreement with the 
the mass shifts of $J/\Psi$, $\Psi (3686)$ and $\Psi (3770)$
as $-8$, $-100$ and $-140$ MeV respectively, at the nuclear matter 
saturation density, computed in Ref. \cite{lee1} from the 
second order QCD Stark effect, with the gluon condensate in the nuclear 
medium computed in the linear density approximation. The 
mass-shift for $J/\psi$ has also been studied with the QCD sum 
rules in \cite{klingl} and the value at nuclear saturation density 
was observed to be about $-7$ MeV. 
In Ref. \cite{kimlee} the operator product expansion was carried out 
upto dimension six and the mass shift for $J/\psi$ was calculated to be 
$-4$ MeV at nuclear matter saturation density $\rho_{0}$ and 
at zero temperature. The effect of temperature on the $J/\psi$ 
in the deconfinement phase was studied in \cite{leetemp,cesa}. 
In these investigations, it was reported that $J/\psi$ mass 
is essentially constant in a wide range of temperatures and 
above a particular value of the temperature, $T$, there was observed
to be a sharp change in the mass of $J/\psi$ in the deconfined phase 
\cite{lee3}. In the present work, we have studied the effects of 
temperature, density and isospin asymmetry, on the mass modifications 
of the charmonium states ($J/\psi, \psi(3686)$ and $\psi(3770)$) 
in the confined hadronic phase, arising due to modifications
of a scalar dilaton field which simulates the gluon condensates of
QCD, within a chiral SU(3) model. The effect of temperature was 
found to be small for the charmonium states $J/\psi(3097)$, 
$\psi(3686)$ and $\psi(3770)$, whereas, the masses of charmonium states 
observed to vary considerably with density, in the present investigation.
 
In summary, in the present work, we have investigated the effects 
of density, temperature and isospin asymmetry of the nuclear medium on the 
masses of the charmonium states $J/\psi, \psi(3686)$ 
and $\psi(3770)$, arising due to modification of the scalar dilaton
field, $\chi$, which simulates the gluon condensates of QCD,  
within the chiral SU(3) model and second order QCD Stark effect. 
The change in the mass of $J/\psi$ with density is observed to be small
at nuclear matter saturation density and is in agreement
with the QCD sum rule calculations. There is seen to be an 
appreciable drop in the in-medium masses of excited charmonium 
states $\psi(3686)$ and $\psi(3770)$ with the density. 
At the nuclear matter saturation density, the mass shifts 
of these states are similar to the values obtained using 
the QCD second order Stark effect with the modifications 
of the gluon condensates computed in the linear density 
approximation \cite{lee1}. 
For a given value of density and temperature, the effect of the 
isospin asymmetry of the medium on the in-medium masses of the 
charmonium states is found to be small. This is due to the fact that
the magnitude of the $\delta$ field remains small as compared to 
the $\sigma$ and $\zeta$ fields. At finite densities, the effect 
of the temperature on the charmonium states is found 
to decrease the values of mass-shift upto a particular temperature,
above which the mass shift is seen to rise. This is because of 
an initial increase in the dilaton field $\chi$ and then a drop
with further increase in the temperature, at a given baryon density, 
arising from solving the coupled equations for the scalar fields.
This is related to the fact that the scalar densities of the nucleons
initially drop and then rise with the temperature at finite values
of the baryon densities. The mass drop of the excited charmonium 
states ($\Psi (3686)$ and $\Psi (3770)$) are large enough 
to be seen in the dilepton spectra emitted from their 
decays in experiments involving $\bar p$-A annihilation
in the future facility at GSI, provided these states decay
inside the nucleus. The life time of the $J/\Psi$ has been
shown to be almost constant in the nuclear medium,
whereas for these excited charmonium states, the lifetimes
are shown to reduce to less than 5 fm/c, due to appreciable
increase in their decay widths \cite{charmwavefn}.
Hence a significant fraction of the produced excited charmonium states
in these experiments are expected to decay inside the nucleus
\cite{Golubeva}. The in-medium properties of the excited charmonium 
states $\psi(3686)$ and $\psi(3770)$ can be studied in the 
dilepton spectra in $\bar p$-A experiments in the future facility 
of the FAIR, GSI. The mass shifts of the charmonium states in the hot
nuclear medium seem to be appreciable at high densities as compared 
to the temperature effects on these masses, and these should show in
observables like the production of these charmonium states, as well as
of the open charmed mesons in the compressed baryonic matter (CBM)
experiment at the future facility at GSI, where baryonic matter 
at high densities and moderate temperatures will be produced.

\acknowledgements
One of the authors (AM) is grateful to the Frankfurt Institute for Advanced 
Research (FIAS), University of Frankfurt, for warm hospitality and 
acknowledges financial support from Alexander von Humboldt Stiftung 
when this work was initiated. Financial support from Department of 
Science and Technology, Government of India (project no. SR/S2/HEP-21/2006) 
is also gratefully acknowledged.


\end{document}